\documentclass[fleqn,usenatbib]{mnras}

\usepackage[T1]{fontenc}

\DeclareRobustCommand{\VAN}[3]{#2}
\let\VANthebibliography\thebibliography
\def\thebibliography{\DeclareRobustCommand{\VAN}[3]{##3}\VANthebibliography}

\usepackage{graphicx}	
\usepackage{amsmath}	
\usepackage{amssymb}	

\usepackage{newtxtext,newtxmath}
\usepackage{makecell}
\usepackage{multirow}

\title[Environments of LAEs at $z$ $\simeq$ 5.7 and $z$ $\simeq$ 6.6]{Can luminous Lyman alpha emitters at $z$ $\simeq$ 5.7 and $z$ $\simeq$ 6.6 suppress star formation?}

\author[D. J. D. Santos et al.]{
Daryl Joe D. Santos,$^{1,2}$\thanks{E-mail: daryl\_santos@gapp.nthu.edu.tw}
Tomotsugu Goto,$^{1}$
Tetsuya Hashimoto,$^{1,3,4}$
Seong Jin Kim,$^{1}$
Ting-Yi Lu,$^{1}$
\newauthor
Yi-Hang Valerie Wong,$^{1}$
Simon C.-C. Ho,$^{1}$
Tiger Y.-Y. Hsiao$^{1}$
\\
$^{1}$Institute of Astronomy, National Tsing Hua University, No. 101, Section 2, Kuang-Fu Road, Hsinchu City 30013, Taiwan \\
$^{2}$Max Planck Institute for Extraterrestrial Physics, Gießenbachstraße 1, 85748 Garching, Germany  \\
$^{3}$Centre for Informatics and Computation in Astronomy (CICA), National Tsing Hua University, 101, Section 2. Kuang-Fu Road, Hsinchu, 30013, \\
Taiwan (R.O.C.)\\
$^{4}$Department of Physics, National Chung Hsing University, 145 Xingda Rd., South Dist., Taichung 40227, Taiwan\\
}

\date{Accepted XXX. Received YYY; in original form ZZZ}

\pubyear{2015}

\begin{document}
\label{firstpage}
\pagerange{\pageref{firstpage}--\pageref{lastpage}}
\maketitle

\begin{abstract}
Addressing how strong UV radiation affects galaxy formation is central to understanding their evolution. The quenching of star formation via strong UV radiation (from starbursts or AGN) has been proposed in various scenes to solve certain astrophysical problems. Around luminous sources, some evidence of decreased star formation has been found but is limited to a handful of individual cases. No direct, conclusive evidence on the actual role of strong UV radiation in quenching star formation has been found.
Here we present statistical evidence of decreased number density of faint (AB magnitude $\geq$ 24.75 mag) Ly$\alpha$ emitters (LAEs) around bright (AB magnitude < 24.75 mag) LAEs even when the radius goes up to 10 pMpc for $z$ $\simeq$ 5.7 LAEs. A similar trend is found for $z$ $\simeq$ 6.6 LAEs but only for faint LAEs within 1 pMpc radius from the bright LAEs. We use a large sample of 1077 (962) LAEs at $z$ $\simeq$ 5.7 ($z$ $\simeq$ 6.6) selected in total areas of 14 (21) deg$^2$ with Subaru/Hyper Suprime-Cam narrow-band data, and thus, the result is of statistical significance for the first time at these high redshift ranges. A simple analytical calculation indicates that the radiation from the central LAE is not enough to suppress LAEs with AB mag $\geq$ 24.75 mag around them, suggesting additional physical mechanisms we are unaware of are at work. Our results clearly show that the environment is at work for the galaxy formation at $z$ $\sim$ 6 in the Universe.

\end{abstract}




\section{Introduction}

Studying the high-redshift (high-$z$) universe allows us to investigate the first formative stages of the Universe. Among these stages, the last one to occur is cosmic reionisation. During this stage, the ultraviolet (UV) and X-ray photons escaping from the first galaxies caused the neutral intergalactic medium (IGM) to be heated and ionised \citep[e.g.,][]{Wise2019}. One of the most important probes of the high-$z$ universe is the hydrogen Lyman alpha (Ly$\alpha$) emission (line). Hydrogen is the most abundant element in the universe, and Ly$\alpha$ emission is produced by hydrogen atoms’ electron transition from n = 2 to n = 1 (ground) state, two of hydrogen’s lowest energy levels. 
\cite{Partridge1967} first predicted that an early galaxy produces powerful Ly$\alpha$ that composes $\sim$6-7\% of its total bolometric luminosity at $z$ $\sim$ 10-30. \cite{Raiter2010} even predicted that the fraction of Ly$\alpha$ that composes the total luminosity may reach up to 20-40\% when the metallicity and the initial mass function (IMF) at higher redshifts are considered. However, it took around 30 years after the first prediction by \cite{Partridge1967} to detect high-$z$ galaxies with prominent Ly$\alpha$ emissions due to the limited sensitivities of the telescopes at the time. 

Galaxies that emit Ly$\alpha$ emission are called Ly$\alpha$ emitters (LAEs). They are believed to be young star-forming galaxies (SFGs) or active galactic nuclei (AGNs) \citep[e.g.,][]{Ouchi2020}, and are compact (with effective radii ranging from $\sim$1-2.5 kpc; e.g., \citealt{Taniguchi2009, Bond2012}), metal-poor (with gas-phase metallicity of around 0.1 ${\rm Z}_\odot$; e.g., \citealt{Nakajima2012, Nakajima2018}), and have low mass (with stellar mass of about $10^8$ – $10^9$ \ ${\rm M}_\odot$; e.g., \citealt{Gawiser2007, Ono2010, Guaita2011, Hagen2014}). Aside from the Ly$\alpha$ emission line’s distinct strength and width at a rest-frame wavelength of 1216Å, the attenuation of Ly$\alpha$ photons due to resonant scattering by neutral hydrogen can put constraints on the ionisation state of the IGM during the first few billion years of the Universe \citep[e.g.,][]{Dayal2009, Rhoads2001, Zheng2017}.

Since the advent of more advanced and more sensitive telescopes, more LAEs are being detected and investigated. LAEs are currently found at various redshift ranges through narrowband (NB) surveys (e.g., with the Subaru/Suprime-Cam, Kitt Peak National Observatory (KPNO), and Hubble Space Telescope (HST) \citep[e.g.,][]{Hu2004, Malhotra2004, Gronwall2007, Konno2014} and spectroscopic observations (e.g., with the FOcal Reducer/low dispersion Spectrograph 2 (FORS2) on the ESO Very Large Telescope (VLT) and the Multi-Object Spectrometer For Infra-Red Exploration (MOSFIRE) on the Keck Telescope \citep[e.g.,][]{Pentericci2011, Finkelstein2013, Schenker2014, Zitrin2015, Stark2017}). However, there are fewer LAEs detected at $z$ $\gtrsim$ 6 compared to lower redshifts. In March 2014, a new large-area NB survey was carried out by the Subaru Strategic Program (SSP; \citealt{Aihara2017}) using the Hyper Suprime-Cam (HSC) on the Subaru telescope. Together with complementary optical and near-infrared (NIR) spectroscopic data, the resulting HSC NB data allowed the detection of an unprecedentedly large sample of LAE candidates at $z$ $\simeq$ 5.7 and 6.6. This created the research project entitled \textit{Systematic Identification of LAEs for Visible Exploration and Reionization Research Using Subaru HSC} (\textit{SILVERRUSH}). Many studies have been already carried out with the \textit{SILVERRUSH} data to get a more accurate picture of cosmic reionisation at higher redshift \citep[e.g.,][]{Ouchi2018, Shibuya2018a, Shibuya2018b, Konno2018, Harikane2018, Inoue2018, Higuchi2019, Harikane2019, Kakuma2021, Ono2021}.

Another area of interest in galaxy formation and evolution is galaxy environment. Many studies have suggested that it is crucial for providing observational insights on structure formation, particularly the hierarchical structure formation scenario. According to this scenario, smaller systems were created first and eventually merged to form larger structures \citep[e.g.,][]{Kauffmann2004}. Understanding the environments of LAEs allows us to learn how they assembled and how their strong star formation (SF) activities started. However, the relationship between the presence of LAEs and galaxy environment is still obfuscated. 


Literature studies have tested environments of quasi-stellar objects/quasars (QSOs) by searching for rest-frame UV bright galaxies with different techniques, mainly detecting Lyman break galaxies (LBGs) with broad-band (BB) filters and LAEs with BB and NB filters. At high redshifts (2 $<$ $z$ $<$ 5), several observations have shown that QSOs have significant overdensities of LAEs in their vicinity. For instance, \cite{Swinbank2012} observed 2 QSOs at $z$ $\sim$ 2.2 and one QSO at $z$ $\sim$ 4.5 using the \textit{Taurus Tunable Filter} (TTF) instrument on the Anglo-Australian Telescope (AAT) with a rectangular field of view (FoV) of 7 $\times$ 5 arcmins. \cite{Husband2013} also targeted the immediate fields of 3 QSOs at $z$ $\sim$ 5, showing that two of these QSOs exhibited significant clustering of LBGs via observations with VLT/FORS2 and HST. \cite{Cantalupo2012} observed a hyperluminous QSO at $z$ = 2.2 using VLT/FORS2, unveiling 98 LAE candidates within a volume of 5500 cMpc$^{3}$ centred at the QSO, while \cite{GarciaVergara2019} targeted 17 QSO fields at $z$ $\sim$ 4 with VLT/FORS2, indicating an LAE auto-correlation length that is about 3 times higher than the value measured in blank fields. This suggests that QSOs at $z$ $\sim$ 4 are within LAE overdensities. However, other studies have presented a different conclusion. \cite{Francis2004} were not able to observe Ly$\alpha$ emission within 1 Mpc of a QSO at $z$ = 2.168 after observing its field with the TTF instrument on AAT. \cite{Kashikawa2007}, on the other hand, implemented NB and BB imaging with Subaru/Suprime-Cam to survey LBGs and LAEs around a QSO at $z$ = 4.87, only to find out that LAEs avoid the immediate vicinity of the QSO up to $\sim$ 4.5 cMpc. Simulations can also be used to investigate this problem, just as \cite{Bruns2012} did, wherein their semi-analytic models suggest that the intense UV emission from the central QSO may have caused the suppression of SF of nearby galaxies, explaining the lack of Ly$\alpha$ emission within a 70 pMpc$^3$ volume centred on a $z$ = 2.168 QSO.

At much higher redshifts ($z$ $>$ 5), studies also show various conclusions. \cite{Djorgovski2003}, using the Palomar 200-inch Hale telescope's prime-focus Cosmic imager, reported a QSO pair at $z$ $\sim$ 5, whose probability of finding one within the given comoving volume ($\sim$ 5-7 $\times$ 10$^4$ Mpc$^3$) is about 10$^{-3}$ to 10$^{-4}$. \cite{Zheng2006} observed a 5 arcmin$^2$ region centred at a radio-loud QSO at $z$ = 5.8 with the HST Advanced Camera for Surveys (ACS) and found that it hosts many $i$-faint objects with a surface density approximately 4-6 times higher than that of the Great Observatories Origins Deep Survey (GOODS) fields. \cite{Overzier2006} and \cite{Venemans2004} discovered 6 LAE companions near a radio galaxy at $z$ = 5.2 with observations from HST/ACS and VLT/FORS2. Furthermore, \cite{Venemans2020} observed 27 QSOs at $z$ $\sim$ 6 with the Atacama Large Millimeter/submillimeter Array (ALMA), and 13 of them have nearby [C II]-emitting galaxies in their fields, with distances from the QSOs ranging from 3 to 88 kpc. Integral-field spectroscopy at higher redshifts with VLT/Multi-Unit Spectroscopic Explorer (MUSE) also reveals the same conclusion. \cite{Meyer2022} discovered 2 LAEs within $<$ 0.6 cMpc on one of the 3 QSOs at 0.2 $<$ $z$ $<$ 6.6 observed by VLT/MUSE. \cite{Farina2017} also detected with VLT/MUSE an LAE $\sim$ 12.5 kpcs away from a $z$ $\sim$ 6.6 QSO. \cite{Bosman2020} found 3 LAEs within the proximity zone of a QSO at $z$ = 5.795 using observations with VLT/MUSE. Cosmological zoom-in simulations from \cite{Zana2022} also support the same conclusion, suggesting that more luminous QSOs' feedback can boost the number density of observable QSO companions at $z$ $\sim$ 6. On the other hand, \cite{Willott2005} undertook deep optical imaging of 3 QSO fields at 6.2 $<$ $z$ $<$ 6.5 with the Gemini Multi-Object Spectrograph North (GMOS-N) and did not see any companion galaxies within them. \cite{Banados2013} employed deep NB and BB imaging of a $z$ = 5.72 QSO with VLT/FORS2, only to find the lack of LAEs within its surroundings. \cite{Mazzucchelli2017} also presented a similar conclusion using the same instrument to observe the 37 arcmin$^2$ field of a QSO at $z$ $\sim$ 5.73. \cite{Goto2017} observed a QSO at $z$ = 6.4 with Subaru/Suprime-Cam, and did not also detect an overdensity of LAEs around it. This lack of LAEs spans up to a large distance of 10 pMpc from the QSO.

There are several reasons for the discrepancy among opposing conclusions. The first one is that the FoVs of the instruments used in previous studies are relatively small (e.g., HST/ACS, which was mostly utilised for studying $z$ $\sim$ 6 QSOs, only has an FoV of 202$''$ $\times$ 202$''$), definitely limiting the region where overdensity around the target could be observed \citep{Goto2017}. Another reason is that previous studies selected galaxies surrounding high-redshift QSOs via the "drop-out technique" based on BB observations. This technique recovers sources with a sharp drop in their continuum emission bluewards of the Ly$\alpha$ emission line wavelength due to the absorption by the neutral intervening intergalactic medium.
This technique also probes a redshift range of $\Delta$$z$ $\approx$ 1. This range is wide enough to identify unwanted high-redshift galaxies that may not be associated with the QSOs (Banados et al. 2013, \citealt{Ota2018}. Previous studies also focused mostly on only one or a few sources for verifying LAE clustering \citep{GarciaVergara2019}.
In addition, most studies that focus on unveiling the environments of high-$z$ quasars are hindered by the uncertainties of the quasar's actual redshifts. Their redshifts are derived from bright rest-frame UV emission lines, which can be strongly shifted with respect to, e.g., the [C II] emission line of the host galaxy, tracing the system rest-frame as recent studies \citep{Meyer2019, Schindler2020} showed. Therefore, the exact redshift of the quasar may be outside of the narrow band's range, and surveys may miss LAEs at the actual redshift of the quasar.

In this work, we focus on understanding the environment of LAEs by utilising the large sample of LAE candidates provided by \textit{SILVERRUSH}. This work is organised as follows: Sec. \ref{sec:data} shows the sample selection and methodology of our work. Sec. \ref{sec:results} shows the main result of our work. Sec. \ref{sec:discussion} presents a discussion about our results, and finally Sec. \ref{sec:conclusion} concludes our work. In our work, we adopt the concordance cosmology with ($\Omega_m$, $\Omega_{\Lambda}$, $h$) = (0.3, 0.7, 0.7) \citep{Planck2016}.

\section{Data Analysis}
\label{sec:data}

\subsection{Sample Selection}
\label{sec:sample_selection}
The full selection of \textit{SILVERRUSH} LAEs is presented in \cite{Shibuya2018a}. Here we report briefly some important details about the selection criteria and the properties of these sources.
The \textit{SILVERRUSH} catalogue version 20171102 \citep{Ouchi2018, Shibuya2018a, Shibuya2018b, Konno2018} is a result of combining HSC SSP S16A data ($g$, $r$, $i$, $z$, and $y$ broadband data; \citealt{Kawanomoto2018}) and NB data (\textit{NB921} and \textit{NB816} data; \citealt{Ouchi2018}). The catalogue contains \textit{NB921} data for 5 ultra-deep (UD) and deep (D) fields (UD-COSMOS, UD-SXDS, D-COSMOS, D-DEEP2-3, D-ELAIS-N1) at z $\simeq$ 6.6, and \textit{NB816} data for 4 fields (UD-COSMOS, UD-SXDS, D-DEEP2-3, D-ELAIS-N1) at $z$ $\simeq$ 5.7. The (area-weighted average) central wavelength and FWHM of the \textit{NB921} (\textit{NB816}) filter are $\lambda_{\rm c}$ = 9215\AA (8177\AA) and $\Delta\lambda$ = 135\AA (113\AA), respectively. In addition, the \textit{NB921} and \textit{NB816} filters can probe the redshifted Ly$\alpha$ emission lines at $z$ = 6.580 $\pm$ 0.056 and $z$ = 5.726 $\pm$ 0.046, respectively. The broadband (BB) filters cover the same area observed by the HSC NB filters and have typical detection limits (5$\sigma$) of $g$ $\simeq$ 26.9, $r$ $\simeq$ 26.5, $i$ $\simeq$ 26.3, $z$ $\simeq$ 25.7, and $y$ $\simeq$ 25.0 ($g$ $\simeq$ 26.6, $r$ $\simeq$ 26.1, $i$ $\simeq$ 25.9, $z$ $\simeq$ 25.2, and $y$ $\simeq$ 24.4) for the UD (D) fields (all within a $1.''5$ aperture). 


Here we present a brief description of the selection criteria utilised \cite{Shibuya2018a}. Sources that have significant flux excess in their NB images and an observed spectral break at the wavelength of redshifted Ly$\alpha$ emission are considered LAE candidates. Final LAE sources are then selected via magnitude and colour selection criteria similar to e.g., \citet{Ouchi2008, Ouchi2010}, excluding blended sources, sources affected by saturated pixels, artificial diffuse objects, cosmic rays, and variable/moving sources.

\textit{SILVERRUSH} LAEs show Ly$\alpha$ equivalent width (EW) distributions that can be explained by exponential and Gaussian distributions. Their best-fitting Ly$\alpha$ EW scale lengths, on average, do not vary between UD and D fields at $z$ $\sim$ 5.7 and 6.6 (see Table 4 of \citealt{Shibuya2018b}), suggesting that these scale lengths do not greatly depend on the image depth or detection completeness. As far as the catalogue’s completeness is concerned, \cite{Konno2018} showed via Monte Carlo simulations that the detection completeness for bright objects (NB $\leq$ 24.5 mag) is $\geq$ 80\%, while at the 5$\sigma$ detection limit, it is $\sim$40\%. The contamination rate is estimated to be 0\% - 30\% at $z$ $\sim$ 5.7 and 6.6 \citep{Shibuya2018b}. Finally, there are 96 spectroscopically-observed LAEs in the catalogue, 21 of them were confirmed by \cite{Shibuya2018b}, while the rest are spectroscopically confirmed by previous studies.

\begin{figure}
	\includegraphics[width=\columnwidth]{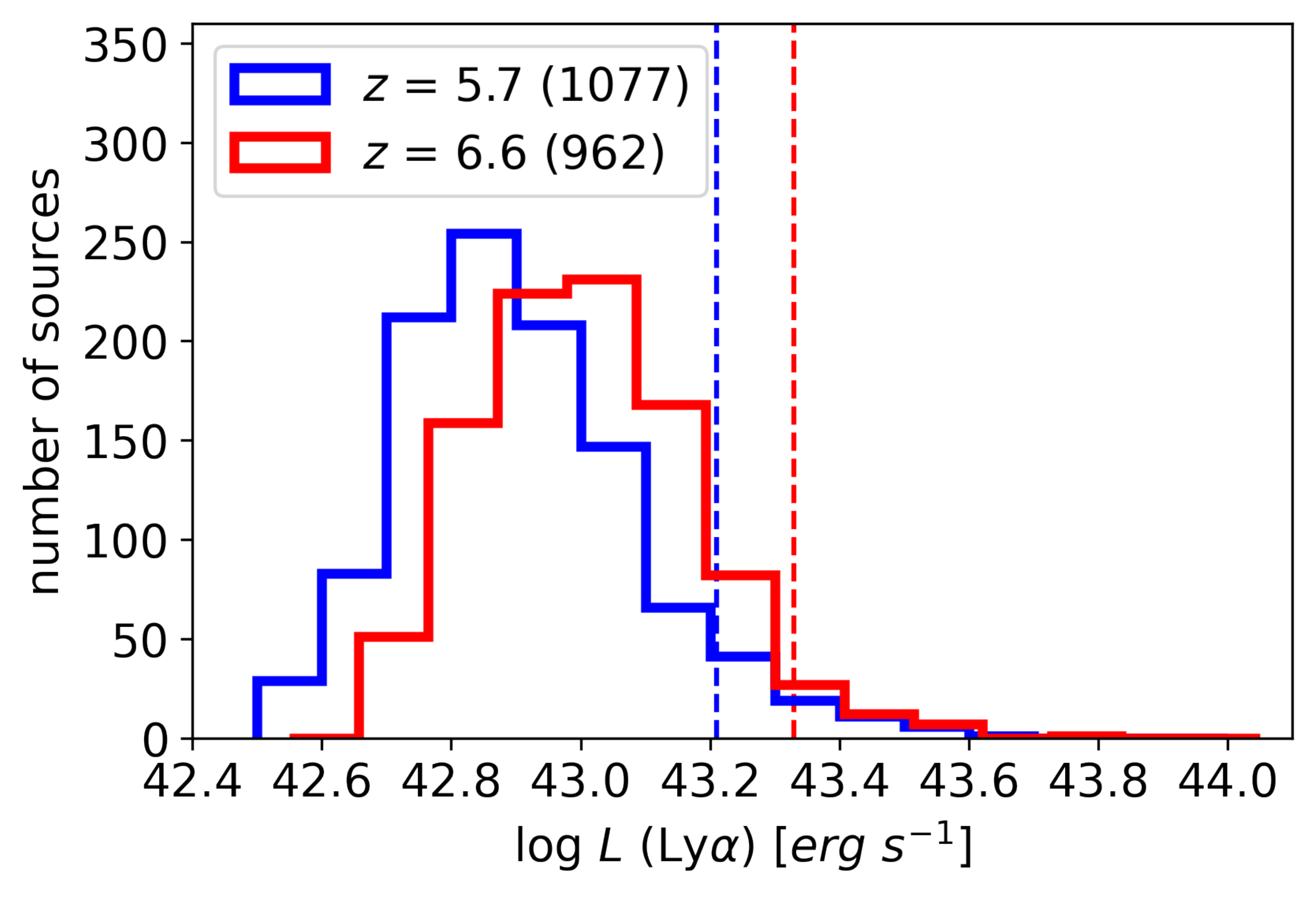}
    \caption{Histogram of Ly$\alpha$ luminosities of our sources. The sources are divided into $z$ $\simeq$ 5.7 (\textit{NB816}) and $z$ $\simeq$ 6.6 (\textit{NB921}). The numbers in the parentheses refer to the total number of sources for each redshift. The coloured dashed lines refer to the luminosity at the given redshifts corresponding to AB mag = 24.75, which is the magnitude cut we used to define faint and bright LAEs (a discussion about our magnitude cut is presented in Sec.~\ref{sec:aperture}. The Ly$\alpha$ luminosities of the sources were calculated from their \textit{NB816} and \textit{NB921} AB magnitudes following our method in Appendix \ref{sec:mag_to_lum}.}
    \label{fig:histogram}
\end{figure}

In our work, we use the \texttt{ALL} catalogue, which is a combination of \texttt{forced} and \texttt{unforced} catalogues. These two catalogues differ in their source detection and photometry in their HSC images. For \texttt{unforced} photometry, one measures the fluxes and coordinates of sources in each band image individually, while in \texttt{forced} photometry, one fixes a centroid determined in a reference band (an NB or BB filter, depending on how bright the sources are in the BB filter) first before measuring the fluxes and coordinates of sources in other band images. Fig. \ref{fig:histogram} shows the histogram of the Ly$\alpha$ luminosities of the sources in the \texttt{ALL} catalogue. A summary of the fields included in our catalogue is presented in Table \ref{tab:sources}.

\begin{table}
\caption{\label{tab:sources}A summary of the properties of the sources in the \textit{SILVERRUSH} \texttt{all} catalogue derived from HSC SSP S16A NB data. The columns show the name of the field, area in deg$^2$, limiting magnitude of the NB image defined by a 5$\sigma$ sky noise in a 1.$^\prime$ $^\prime$ diameter circular aperture, and the number of LAEs, ${\rm N}_{\rm LAE, ALL}$.}
\begin{center}
\begin{tabular}{cccc}
\hline
Field & Area (deg$^2$) & \makecell{Depth \\ (5$\sigma$ AB mag)} & ${\rm N}_{\rm LAE, ALL}$ \\
\hline
\multicolumn{4}{c}{\textit{NB816} ($z$ $\simeq$ 5.7)} \\
\hline
UD-COSMOS & 1.97 & 25.7 & 201\\
UD-SXDS & 1.93 & 25.5 & 224\\
D-DEEP2-3 & 4.37 & 25.2 & 423\\
D-ELAIS-N1 & 5.56 & 25.3 & 229\\
\textbf{Total} & \textbf{13.8} & -- & \textbf{1077}\\
\hline
\multicolumn{4}{c}{\textit{NB921} ($z$ $\simeq$ 6.6)} \\
\hline
UD-COSMOS & 2.05 & 25.6 & 147$^{a}$\\
UD-SXDS & 2.02 & 25.5 & 58\\
D-COSMOS & 5.31 & 25.3 & 244$^{b}$\\
D-DEEP2-3 & 5.76 & 24.9 & 164\\
D-ELAIS-N1 & 6.08 & 25.3 & 349\\
\textbf{Total} & \textbf{21.2} & -- & \textbf{962}\\
\hline
\end{tabular}
\end{center}
$^{a}$Very faint sources (i.e., sources fainter than 5$\sigma$ limiting magnitude) from the NB921 UD-COSMOS catalogue were removed. This criterion makes no or negligibly small impact on the main scientific results of our work.\\
$^{b}$30 LAEs selected in NB921 UD-COSMOS were also included in the NB921 D-COSMOS catalogue.
\end{table}

\subsection{Aperture Counts, Edge Correction, and Magnitude Cut}
\label{sec:aperture}

In this work, we analyze the number of "faint" LAEs around "bright" LAEs within a certain annulus size using 1077 (962) sources at $z$ $\simeq$ 5.7 ($z$ $\simeq$ 6.6) detected by HSC \textit{NB816} (\textit{NB921}) bands. Many studies, both observational and theoretical ones, have suggested that small-scale environments and dark matter halo mass, which can be probed by aperture counts, have more influence in galaxy properties compared to large-scale environments \citep[e.g.,][]{Lemson1999, Blanton2006}. However, due to the redshift uncertainty, the scale with respect to the line of sight is larger than the aperture size. If the observed suppression is only evident within a certain radial distance or aperture size from the bright LAEs, we may see that within this radial distance, the density of faint LAEs is lower. But as the aperture size increases, noise is further added and so the low-density environment is less clear. Hence, we decided to investigate the density of faint LAEs within annuli instead of apertures.

To further understand whether surrounding faint LAEs are suppressed by their bright central galaxies, we introduce a magnitude cut to define "bright" and "faint" LAEs. In our work, "bright" ("faint") LAEs are defined as those with \textit{NB816}/\textit{NB921} AB mag $<$ ($\geq$) 24.75, which translates to log $L$ (Ly$\alpha$) [erg s$^{-1}$] $\sim$ 43. This magnitude cut gives us 605 (566) faint LAEs to count and 472 (396) bright LAEs at $z$ $\simeq$ 5.7 ($z$ $\simeq$ 6.6) to consider as central LAEs. Therefore, central LAEs are always brighter than the faint LAEs we are counting. \cite{Shibuya2018b} used a similar magnitude cut to select bright LAEs for follow-up spectroscopic observations. \cite{Konno2018} studied the luminosity function derived from SILVERRUSH LAEs, showing a characteristic Ly$\alpha$ luminosity (L*) of L* = 1.6$^{-0.6}_{+2.2}$ $\times$ 10$^{43}$ erg/s and L* = 1.7$^{-0.7}_{+0.3}$ $\times$ 10$^{43}$ erg/s for $z$ $\simeq$ 5.7 and $z$ $\simeq$ 6.6, respectively. Our choice of magnitude cut corresponds to luminosities that are quite fainter than L* for both redshifts. We also tried varying the magnitude cut by adding/subtracting AB mag = 0.25 in the cut, but our results do not drastically change in terms of the corresponding luminosity and number of faint and bright LAEs at both redshifts, so we conclude that the magnitude cut does not greatly affect our results. The choice of the magnitude cut was also made to make sure that there are enough number of faint LAEs to count around central LAEs and enough number of bright LAEs to consider as central LAEs.

Here we briefly compare our selected faint LAEs to the LAEs from works focusing on similar redshifts that found LAE overdensities around luminous QSOs. One difference is that our work utilises an order of 3 magnitudes higher number of LAEs compared to those from previous works, which only focused on a very few numbers of LAEs and QSOs. Previous works \citep{Meyer2022, Bosman2020, Farina2017} have discovered LAEs with log $L$ (Ly$\alpha$) $\leq$ 43, which are quite similar to the corresponding luminosities for our chosen magnitude cut, which corresponds to log $L$ (Ly$\alpha$) $\sim$ 42.91 and $\sim$ 43.03 for $z$ $\simeq$ 5.7 and 6.6, respectively. \cite{Bosman2020} also discovered that one of the three LAEs surrounding their target QSO is a double-peak Ly$\alpha$ emitter, which we do not have in our sample.


\cite{Konno2016} previously showed Ly$\alpha$ luminosity functions (LFs) at $z$ = 2.2 have an excess beyond log $L$ (Ly$\alpha$) [erg ${\rm s}^{-1}$] $\gtrsim$ 43.4. By studying the accompanying multiwavelength data, they believe that this excess is caused by (faint) AGNs. Other studies have also found bright LAEs at redshifts higher than 2.2 \citep[e.g.,][]{Ouchi2008, Sobral2015}. \cite{Konno2018} also found similar results using LAEs from \textit{SILVERRUSH} wherein the steeper slope of the LAE LF starting at log $L$ (Ly$\alpha$) [erg ${\rm s}^{-1}$] $\gtrsim$ 43.5 might be due to AGNs. However, \cite{Shibuya2018b} revealed via follow-up spectroscopic observations of LAEs with log $L$ (Ly$\alpha$) [erg ${\rm s}^{-1}$] $\gtrsim$ 43.5 (equal to $\sim$23.5 AB mag) at $z$ = 5.7 and $z$ = 6.6 from \textit{SILVERRUSH} that most of these LAEs lack AGN signatures (i.e., broad Ly$\alpha$ emission line, strong highly ionised metal lines, etc.). Even though there are 96 spectroscopically observed \textit{SILVERRUSH} LAEs, more follow-up deep NIR spectroscopic observations must be carried out to confirm the identities of these bright sources.

After defining the bright and faint LAEs, we define the annulus sizes ranging from $<$ 1.0 pMpc, 1.0 - 2.0 pMpc, 2.0 - 3.0 pMpc, 3.0 - 5.0 pMpc, and 5.0 - 10.0 pMpc. These annulus sizes accommodate different scales of galaxy environment, which range from individual dark matter haloes ($\sim$1 pMpc) to large voids in the cosmic web ($\sim$10 pMpc). For each bright (central) LAE, we count the number of faint LAEs within a radius defined by the annulus size. The LAE density (number count within a luminosity bin divided by the area of the annulus in pMpc$^{-2}$) of the bright central LAE is then calculated.

\begin{figure}
	\includegraphics[width=\columnwidth]{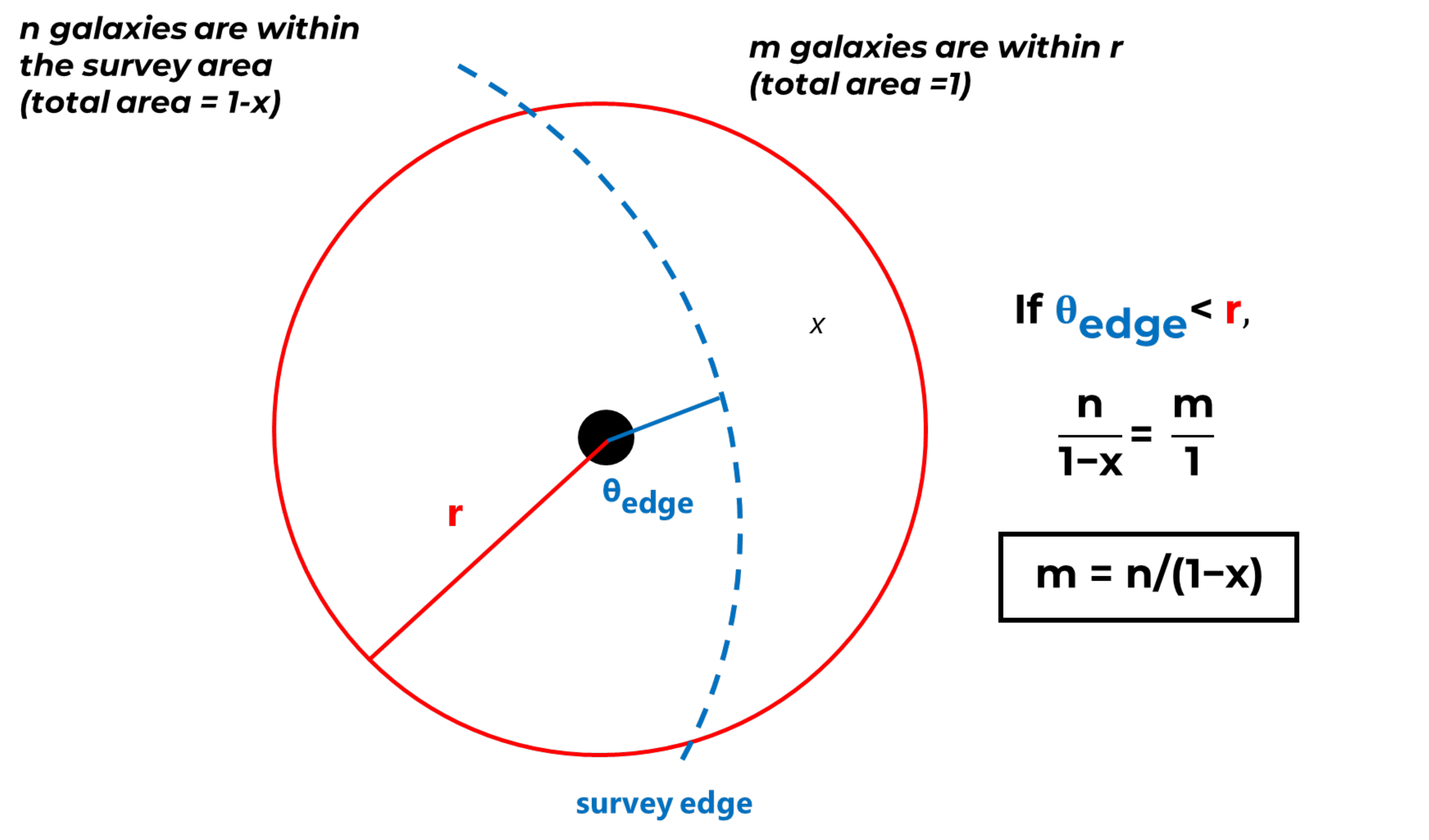}
    \caption{A schematic diagram showing the edge correction method used in this work. The solid black circle is the target source, $r$ is the aperture size, and $\theta_{\rm edge}$ is the distance of the source to the survey edge, which is denoted by the blue dashed line. Within the red circle traced by the aperture size, $r$, there should be {\tt m} galaxies. However, due to the limited survey area, we can only see {\tt n} galaxies within the circle. With our edge correction method, we can estimate {\tt m} using the given formula {\tt m = n/(1-x)}, where {\tt x} is the approximate fractional area of the red circle that is not covered by the survey.}
    \label{fig:edge_correction}
\end{figure}

Aperture counts suffer from edge effects. This happens when a source is near the survey edge, and the distance of the source from the survey edge is less than the aperture size. This causes the other galaxies not covered by the survey to be missed in the count, causing the measured aperture count to be smaller than what it should be. Due to edge effects, previous studies \citep[e.g.,][]{Miller2003, Cooper2005} removed these edge sources in their analyses. \cite{Santos2021} proposed an edge correction method to estimate the correct local galaxy density by scaling the measured counts within a given radius by the amount of aperture area covered by the survey. We use the same edge correction method revised for aperture counts to measure the environment of our edge sources. Fig. \ref{fig:edge_correction} shows a visualisation of our edge correction method. A circular area with the aperture size, $r$, as the radius is created, and we let {\tt x} as the approximate area that is not covered by the survey. If the edge distance, $\theta_{\rm edge}$, is less than $r$, then the true number of galaxies within $r$, which we regard as {\tt m} is estimated by the following formula:

\begin{eqnarray}
    {\tt m = \frac{n}{1-x}}
    \label{eq:edgecorrection}
\end{eqnarray}

where {\tt n} is the number of galaxies within $r$ that is covered by the survey, and 1-{\tt x} is the percent/fractional area that is covered by the survey. With edge correction, we expect that the true number of faint galaxies within the given aperture size will be larger than that without edge correction since we are taking into consideration the galaxies not covered by the survey. Our edge correction method is done under the assumption that galaxy densities are the same in and outside of the edge within the aperture. This also entails that increasing the aperture size will also increase the uncertainty or noise added to the edge-corrected density \citep{Santos2021}. This correction would overestimate the density if the area outside of the edge is a lower density region (e.g. an edge of the cluster). On the contrary, it might underestimate if the area outside of the edge has a secondary density peak. In any case, the area we correct for the density is always less than 50\%. More details about the robustness of the edge correction method are discussed in \citet{Santos2021}. Table \ref{tab:edge_correction} shows the number of edge galaxies (bright central LAEs whose aperture counts are estimated by edge correction) in \textit{SILVERRUSH} \textit{NB816} and \textit{NB921} \texttt{all} catalogues. As expected, increasing aperture size causes the number of edge galaxies to increase. Therefore, we caution the readers that edge correction is affecting a significant amount of galaxies at larger aperture sizes, and so the relative results at these sizes should be taken with care.

\begin{table}
\caption{\label{tab:edge_correction} Number of bright central LAEs with edge-corrected densities (edge galaxies) in \textit{NB816} and \textit{NB912} \textit{SILVERRUSH} catalogues in each aperture size. Columns refer to the aperture size in pMpc, the number of edge galaxies, and percentage of edge galaxies.}
\begin{center}
\begin{tabular}{ccc}
\hline
r (pMpc) & \makecell{Number of \\ edge galaxies} & \makecell{Percentage of \\ edge galaxies} \\ 
\hline
\multicolumn{3}{c}{\textit{NB816} ($z$ $\simeq$ 5.7)}\\
\hline
1 & 15 & 3.2\% \\
2 & 52 & 11.0\% \\
3 & 93 & 19.7\% \\
5 & 162 & 34.3\% \\
10 & 330 & 69.9\% \\
\hline
\multicolumn{3}{c}{\textit{NB921} ($z$ $\simeq$ 6.6)}\\
\hline
1 & 11 & 2.8\% \\
2 & 60 & 15.2\% \\
3 & 92 & 23.2\% \\
5 & 166 & 41.9\% \\
10 & 311 & 78.5\% \\
\hline
\end{tabular}
\end{center}
\end{table}

An annulus’s size is defined by a smaller radius ${\tt r_1}$ and a larger radius ${\tt r_2}$, both in pMpc. The number of faint LAEs within an annulus is defined as the aperture counts in ${\tt r_1}$ subtracted from the edge-corrected aperture counts in ${\tt r_2}$. This way, we can utilize edge-corrected annulus counts for our analyses.

\section{Results}
\label{sec:results}

\begin{figure*}
	\includegraphics[width=\textwidth]{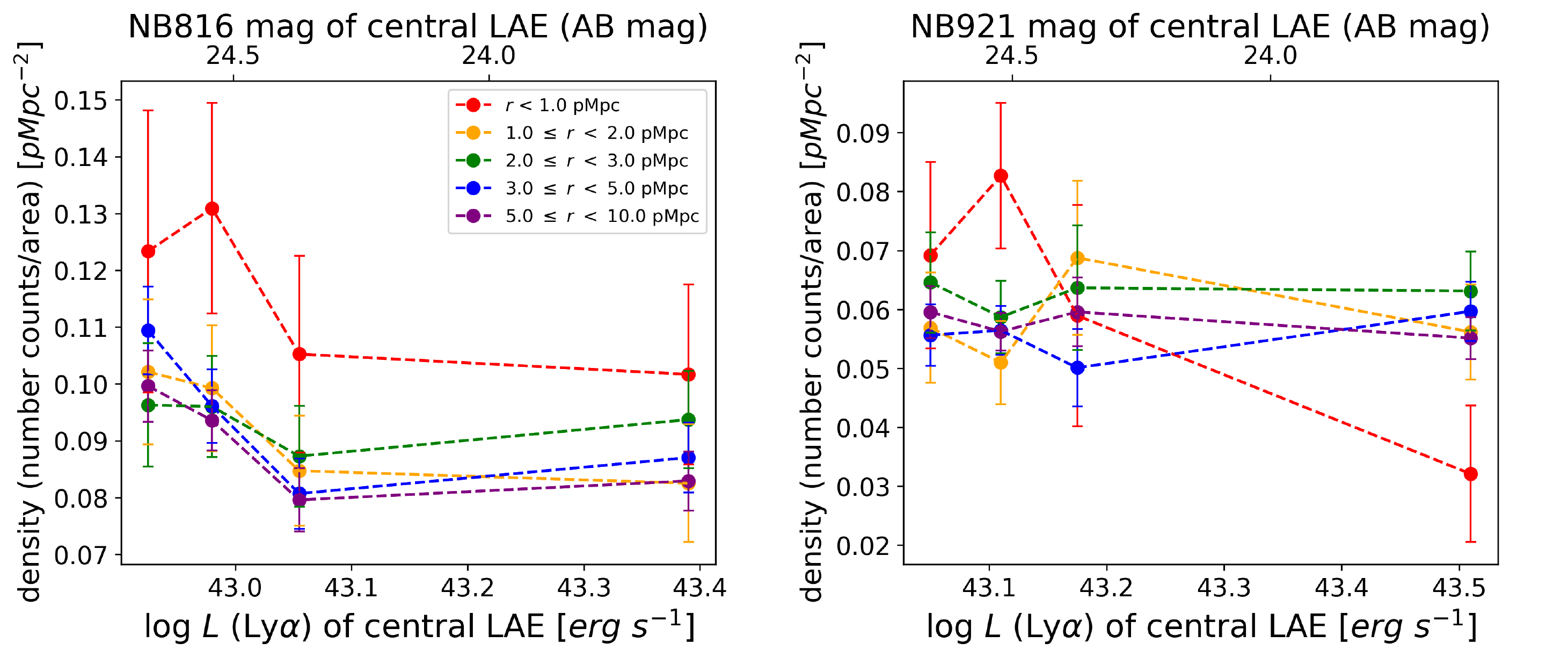}
    \caption{The density of faint LAEs (number of faint LAEs within a luminosity bin divided by the area of the annulus, in pMpc$^{-2}$) as a function of logarithm of Ly$\alpha$ luminosityof the bright central LAEs at \textit{NB816} ($z$ $\simeq$ 5.7) and \textit{NB921} ($z$ $\simeq$ 6.6) (left and right panel, respectively). The top horizontal axis refers to the AB magnitudes corresponding to the Ly$\alpha$ luminosities of the central LAEs. We use all LAEs fainter than AB mag = 24.75 to measure the aperture counts. Error bars correspond to standard errors. Different colours pertain to different annulus sizes as mentioned in the legends. The sources are binned in terms of Ly$\alpha$ luminosity so that each bin has at least 50 sources.}
    \label{fig:density_vs_luminosity}
\end{figure*}

Fig. \ref{fig:density_vs_luminosity} shows the density of faint LAEs (LAEs with AB magnitude $\geq$ 24.75 mag) as a function of the logarithm of Ly$\alpha$ luminosity of the central LAE for $z$ $\simeq$ 5.7 and $z$ $\simeq$ 6.6. To create Fig. \ref{fig:density_vs_luminosity}, we first count the number of faint LAEs (LAEs with AB magnitude $\geq$ 24.75) around all bright LAEs within a certain annulus size (with edge correction whenever necessary). The density of faint LAEs is then calculated for each bright LAE. After that, the bright LAEs are binned in luminosity bins so that each luminosity bin has at least 50 LAEs. In each luminosity bin, the average density and standard error bars are calculated. The luminosity values plotted in Fig. \ref{fig:density_vs_luminosity} are the midpoints of our luminosity bins. Different line colours pertain to different annulus sizes. 

For the bright LAEs at $z$ $\simeq$ 5.7, it is clear that the density of faint LAEs surrounding them decreases as their Ly$\alpha$ luminosity increases. Both annulus sizes show a $\sim$20\% decrease of density at the highest luminosity bin compared to the value at the lowest luminosity bin except for the annulus between 2.0 and 3.0 pMpc which showed less degree of decrease ($\sim$ 2\% only) (see Appendix \ref{sec:densities}). However, the bright LAEs at $z$ $\simeq$ 6.6 do not show a similar trend for all annuli except the smallest aperture (< 1.0 pMpc). For radii > 1.0 Mpc, the density of faint LAEs around bright LAEs is almost constant. Another trend is also apparent at $z$ $\simeq$ 5.7, wherein as the annulus size increases, the density of faint LAEs around central LAEs decreases regardless of the central LAEs' luminosities. However, this trend is not clear at $z$ $\simeq$ 6.6. We believe that the lack of trends at $z$ $\simeq$ 6.6 is due to the observability of LAEs at high redshifts, when the epoch of reionisation is at play (see Sec.~\ref{sec:other_reasons}), or simply due to poorer statistics at this redshift.


Nevertheless, our results show that the number of faint LAEs at smaller apertures (as low as 1.0 pMpc) is not zero for both redshift ranges, indicating that the observed suppression is not capable of completely destroying faint LAEs. One possibility is that these galaxies have collapsed outside of the halo, later to be accumulated into the halo. Or more simply, there may be a projection effect in that actual distance in the line of sight direction which is much larger than the angular separation.

\section{Discussion}
\label{sec:discussion}

\subsection{Estimation of background UV radiation strength from central LAEs}
\label{sec:estimation_uv}

Our results show that the density of LAEs with AB magnitude $\geq$ 24.75 mag around central LAE decreases as the luminosity of the central LAE increases. This trend is apparent for $z$ $\simeq$ 5.7 regardless of the annulus sizes from the bright LAEs. A similar trend is also apparent for $z$ $\simeq$ 6.6 but only at distances < 1.0 Mpc from the bright LAEs.
One plausible reason for this trend is that central LAEs' ionising photons suppress the star formation of surrounding galaxies and thus reduce the number of LAEs around them. Hence, we quantitatively estimate the strength of UV radiation from our bright central LAEs to verify this. Since we do not have NB observations for the continuum flux at the Lyman limit (912\r{A} in rest-frame, 6110\r{A} for the sources at $z$ $\simeq$ 5.7, 6930\r{A} as observed in $z$ $\simeq$ 6.6), we use the upper limit of \textit{SILVERRUSH} $i$-band ($\lambda_{\rm eff} = 7711$\r{A}, $\Delta\lambda$ = 1574\r{A}) observations to estimate the upper limit of ionising photons' radiation strength. The $i$-band limiting magnitude of our sources is $i$ $\simeq$ 26.3 and $i$ $\simeq$ 25.9 for UD and D fields, respectively \citep{Shibuya2018a}. 882/1077 (705/962) of the sources detected in \textit{NB816} (\textit{NB921}) are not detected in the $i$-band. To calculate the upper limit of the UV radiation from central LAEs, we choose $i$ $\simeq$ 26.3 for our calculations. The flux density at the Lyman limit ($f_{\nu}$), can be calculated using the following formula \citep{Fan2001}:

\begin{eqnarray}
    m_{\rm AB} = -2.5 {\rm log}_{\rm 10} f_{\nu} - 48.60,
	\label{eq:mag_to_flux}
\end{eqnarray}

where $m_{\rm AB}$ is the AB magnitude (in this case, we are substituting the limiting magnitude of our sources at $i$-band). Substituting the value of $f_{\nu}$ in units of erg s$^{-1}$ cm$^{-2}$ Hz$^{-1}$, we can calculate the luminosity of the central LAEs at the Lyman limit \citep{Calverley2011}:

\begin{eqnarray}
    {\rm L}_\nu = 4\pi d^{2}_{L} \frac{f_{\nu}}{(1+z)},
	\label{eq:lum}
\end{eqnarray}

where $d_{L}$ is the luminosity distance at the redshift of the source, $z$, and ${\rm L}_\nu$ is in units of erg s$^{-1}$ Hz$^{-1}$. For a given distance from the source ($r$), the Lyman limit flux density is given by \citep{Calverley2011}:

\begin{eqnarray}
    F_{\nu}(r) = \frac{{\rm L}_\nu}{4\pi r^{2}},
	\label{eq:density}
\end{eqnarray}

In terms of UV intensity, we have $J_\nu$ = $F_{\nu}(r)/4\pi$ where $J_\nu$ is in units of erg s$^{-1}$ cm$^{-2}$ Hz$^{-1}$ ster$^{-1}$. In terms of isotropic UV intensity at the Lyman limit, $J_{21}$, we have the following formula \citep{Calverley2011}:

\begin{eqnarray}
    J_\nu = J_{21} \left( \frac{\nu}{\nu_L} \right)^{\alpha} \times 10^{-21} \ {\rm erg} \ {\rm s}^{-1} \ {\rm Hz}^{-1} \ {\rm cm}^{-2} \ {\rm ster}^{-1},
	\label{eq:intensity}
\end{eqnarray}

where we use the effective frequency of HSC $i$-band for $\nu$, and the expected observed frequency of Ly$\alpha$ limit for $\nu_{L}$. We used $\alpha$ = -1 in our work, similar to that of \cite{Kashikawa2007}. However, this value could be different from the expected $\alpha$ for the \textit{SILVERRUSH} LAEs since \cite{Kashikawa2007} used this value which is close to their observed LAE at $z$ = 4.87.

\begin{figure}
	\includegraphics[width=\columnwidth]{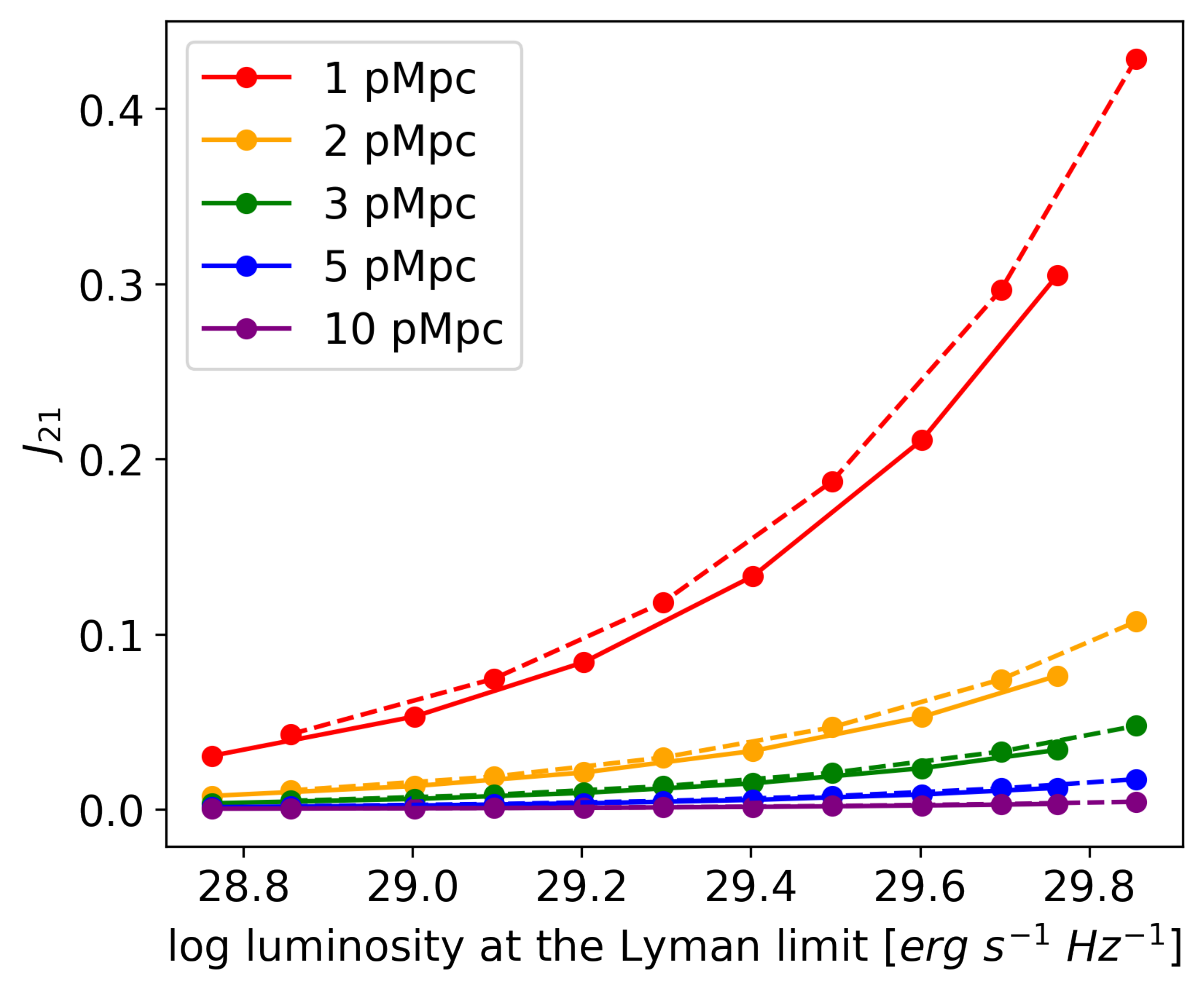}
    \caption{(Left panel) Estimated $J_{21}$ values of our sources as a function of logarithm of luminosity at the Lyman limit of a bright central LAE. The $J_{21}$ values are calculated using magnitudes ranging from the $i$-band limiting magnitude of our observations (26.3) up to the smallest $i$-band magnitude for our LAEs ($i$ $\sim$ 23.8 and $i$ $\sim$ 24.6 for $z$ $\simeq$ 5.7 and $z$ $\simeq$ 6.6 LAEs, respectively). Different colours pertain to different aperture sizes/distances from the central LAEs. Solid (dashed) lines refer to $J_{21}$ values estimated at $z$ $\simeq$ 5.7 ($z$ $\simeq$ 6.6).}
    \label{fig:j21}
\end{figure}

Fig. \ref{fig:j21} shows $J_{21}$ as a function of the logarithm of luminosity at the Lyman limit. This shows the possible strength of the UV background radiation from the central LAEs in our work. At $z$ $\simeq$ 6.6, the highest $J_{21}$ that we can estimate from our data is $\sim$0.43 (with $r$ = 1 pMpc). On the other hand, the highest $J_{21}$ that we can estimate from our data at $z$ $\simeq$ 5.7 is $\sim$0.30. Our $J_{21}$ estimates span from 0.43 to $10^{-4}$ at $r$ = 10 pMpc for both redshifts. For comparison, \cite{Goto2017} estimated $J_{21}$ $\sim$ 24 within 770 kpc from a QSO at $z$ = 6.4 whose absolute magnitude at 1450\r{A} is $M_{\rm 1450\text{\r{A}}} = -25.2$. Their work used an absolute magnitude at 1450\AA \ which lies within the $z$-band. The large difference in our estimations with that of \cite{Goto2017}'s could be attributed to the brightness and distance difference between our sources and their source. For instance, \cite{Goto2017}'s QSO has a $z$-band magnitude of 21.165 mag, while the upper limit of $z$-band magnitude of our sources is 25.7 mag. This translates to a 4 order of magnitude difference and so our sources are a factor of 100 fainter. Therefore, we expect that our $J_{21}$ at $r$ = 1 pMpc will be 2 orders of magnitude weaker compared to what \cite{Goto2017} estimated within 770 kpc from their QSO. As for our estimated $J_{21}$ at $r$ = 10 pMpc, the distance difference will also induce a decrease in the estimated $J_{21}$. $r$ = 10 pMpc is almost 1 order larger compared to $r$ = 770 kpc from \cite{Goto2017}'s work. Since $J_{21}$ is inversely proportional to the square of the distance, this means that our estimated $J_{21}$ will be 2 orders of magnitude weaker. Adding the brightness difference will result in a total of 4 orders of magnitude weaker $J_{21}$ at $r$ = 10 pMpc. These factor reductions are clearly observed in our estimates.



Our $J_{21}$ estimates suggest that the background UV radiation is stronger at smaller radii, which can explain the suggested higher efficiency of suppression of faint LAEs at smaller distances from the central LAEs, and the lack of decreasing trend between faint LAE density and Ly$\alpha$ luminosity of bright central LAEs at $z$ $\simeq$ 6.6 at distances > 1.0 pMpc. Although our estimated $J_{21}$ are within our expectations, the estimated value of $J_{21}$ in our work is not sufficient to explain the decreasing trend in the number counts of faint LAEs as a function of Ly$\alpha$ luminosity.

\cite{Kashikawa2007} showed the relationship between the time delay in SF of the sources within the vicinity of a QSO and its $J_{21}$ with varying halo masses of the LAEs (see Fig. 7 in their work). According to their work, star formation in haloes with $M_{\rm vir}$ < $10^{9} \ {\rm M}_\odot$ is completely suppressed when exposed to a background UV field with $J_{21}$ $\sim$ 1. To achieve this conclusion, \cite{Kashikawa2007} they used a radiation-hydrodynamic simulation model that is limited to star formation at the centre of a nearly spherical halo. This may cause uncertainties when star formation can take place due to disk-like collapse of pre-galactic clouds \citep{Susa2000} and/or collapse of smaller substructures of gas and dust into larger systems \citep{Mori2006}. As for the halo masses of our sources, \cite{Ouchi2018} selected 959 and 873 \textit{SILVERRUSH} LAEs at $z$ $\simeq$ 5.7 and 6.6, respectively, to measure their average and minimum halo masses. These homogeneous LAEs are selected so that they are much brighter than the 5$\sigma$ detection in NB. With halo occupation distribution (HOD) modeling, the minimum halo masses of \textit{SILVERRUSH} LAEs are estimated to be around $M_{\rm vir} \sim 10^{9}$ \ ${\rm M}_\odot$, while their average halo masses range from $M_{\rm vir} \sim 10^{10} - 10^{11} \ {\rm M}_\odot$. 
Considering the halo masses of \textit{SILVERRUSH} LAEs and the time delay-$J_{21}$ relation presented by \cite{Kashikawa2007}, our $J_{21}$ estimates show that faint LAEs with $M_{\rm vir}$ < $10^{9} \ {\rm M}_\odot$ can have their SF activities hindered by the radiation strength from nearby LAEs. However, since the average halo masses of our \textit{SILVERRUSH} LAEs are at approximately $M_{\rm vir} \sim 10^{10} - 10^{11} \ {\rm M}_\odot$, this cannot explain the SF suppression of faint \textit{SILVERRUSH} LAEs with $10^{9} \leq M_{\rm vir} < 10^{11}$.


\subsection{Other possible reasons for decreasing trend between the density of faint LAEs and luminosity of central LAEs}
\label{sec:other_reasons}
For the first time, we can shed light on the growing tension between opposing results of earlier works studying the environments of LAEs and QSOs by statistically analysing an unprecedentedly large sample of LAEs instead of focusing on just one source which previous studies did. Our results suggest two things:
\begin{enumerate}
    \item Suppression of faint LAEs around brighter LAEs is more efficient at smaller radial distances from brighter LAEs (which is true for our $z$ $\simeq$ 5.7);
    \item At $z$ $\simeq$ 5.7, more luminous LAEs prefer less dense environments of faint LAEs from up to 10 pMpc from them; and
    \item At $z$ $\simeq$ 6.6, more luminous LAEs prefer less dense environments of faint LAEs within a distance 1.0 pMpc
\end{enumerate}

Our first and third conclusion can be explained by higher J$_{21}$ values at smaller distances from the central LAEs, as shown in Sec.~\ref{sec:estimation_uv}, should this be accepted at its face value. However, as seen in Fig.~\ref{fig:density_vs_luminosity}, only at our $z$ $\simeq$ 5.7 sample we see the trend between density of faint LAEs and annulus size. For our sample at $z$ $\simeq$ 6.6, we do not clearly observe the density of faint LAEs around central LAEs to increase with decreasing annulus size. The presence of this trend in our $z$ $\simeq$ 5.7 sample might be due to the visibility of LAEs at the epoch of reionisation in play. LAEs create ionised bubbles around them, which may span a few pMpc \citep{Endsley2022}. Ly$\alpha$ photons are opaque to neutral IGM, so they prefer ionised regions to propagate. If we have a relatively faint central LAE, as we probe smaller annulus sizes, we expect to find more faint LAEs. However, as we probe larger annulus sizes, we start to probe the regions outside the faint central LAE's ionised bubble, which will cause the observed density of faint LAEs to be smaller. However, if we consider the central LAE to be relatively brighter, we also expect its ionised bubble to be larger \citep{Weinberger2018}. Therefore, we still expect to see the increased density of faint LAEs at smaller annulus sizes, but its decrease towards larger annulus sizes will be less drastic. Note that this does not take into account the increase of ionised bubble size as the clustering of LAE increases \citep{Yajima2018}. However, this follows that more luminous LAEs with larger ionised bubbles must also have increased faint LAE number density. This does not agree with our results, as they show that luminous LAEs have fewer faint LAEs around them. This suggests that ionised bubbles are not sufficient to explain our results. Another possibility is that the density profiles of our sources peak at the centre/near the central LAE, and fall down at larger distances. But this is not enough to explain why as the central LAE luminosity increases, the decrease in faint LAE density as annulus size increases is less drastic. Future work in this direction will be beneficial in understanding the underlying physics behind this phenomenon.
On the other hand, the absence of this trend in our $z$ $\simeq$ 6.6 sample could be due to its smaller sample size compared to that at $z$ $\simeq$ 5.7, which could be simply due to the larger luminosity distance, added by the effect of a more neutral IGM at $z$ $\simeq$ 6.6 that affects the visibility of LAEs at much higher redshift. The effect of IGM in our work is discussed in a later part of this section. This could also cause the lack of decreasing trend between density of faint LAEs and Ly$\alpha$ luminosity of bright LAEs at $z$ $\simeq$ 6.6.






Meanwhile, our the second conclusion is similar to those from works that show the lack of surrounding galaxies/LAEs around QSOs \citep[e.g.,][]{Willott2005,  Banados2013, Goto2017, Mazzucchelli2017}, except that we used order of 3 larger magnitude of LAEs instead of a few sources. However, our bright LAEs are most likely to be non-AGNs \citep{Shibuya2018b}, and as we showed in our simple estimation in Sec.~\ref{sec:estimation_uv}, the estimated background UV radiation provided by the bright LAEs is not enough to explain the observed trend in our work. Therefore, it is important to look into other tenable explanations for this phenomenon.

The apparent decrease in the number density of faint LAEs for all possible halo masses as a function of Ly$\alpha$ luminosity  may be explained by certain reasons. 
QSO activities near the bright LAEs might have contributed most of the ionising photons, but these quasars might have an episodic or very short period of luminous phase, rendering them undetectable by the time they were observed \citep{Wyithe2005, Vanzella2011, Martini2004}. In addition, there is still an uncertainty in the halo masses of LAEs since they depend on the age of the stellar population \citep{Goto2017}. Further research to confirm the true nature of the bright and faint LAEs in our sample would be beneficial to explain the consistent trends in our work. For instance, ultradeep spectroscopy may be able to detect high-ionization emission lines caused by AGN, or other possible contaminants in the sample which may show other emission lines \citep{Stanway2007}. Deep multiwavelength photometric observations, on the other hand, may also help in having a more extensive characterization of these sources by providing spectral energy distributions of faint and bright SILVERRUSH LAEs \citep[e.g.,][]{Harikane2018}. In addition, galaxies move in real-time, i.e., our simple evaluation of the radiation field is based on the current positions of the LAEs as they were observed. Since galaxies move and mix during the formation process of their haloes, some suppression effects are expected to propagate to larger radii. There is a lack of studies unveiling the possible velocity dispersions of LAEs at very high redshifts. Galaxies can have velocity dispersions ranging from orders of $\sim$ 10$^1$ - 10$^3$ km/s \citep{Struble1999}. Assuming a velocity dispersion of about 10$^3$ km/s, a galaxy will take $\sim$1 Gyr to move 1 Mpc away from its original position. On the other hand, the star formation timescale (t$_{SF}$) can be defined as the average time needed for molecular clouds to form massive stars. A simple estimate of t$_{SF}$ could be derived from the Jeans timescale of molecular clouds, which is in order of approximately 10$^7$ years \citep{Egusa2004}. As star formation happens first, these galaxies tend to use up their fuel before moving 1 Mpc away from their original position. However, these are crude estimates without regard to the redshift of our sources. Since we were able to observe a similar trend up to 10 Mpc, this would mean that our sources must have a much higher velocity dispersion than our assumed value. Another point to be considered is that in the case that such high velocities are involved, some of them might fall outside the NB filter (e.g., $\geq$ 4100 km/s for NB816 filter, and $\geq$ 4400 km/s for NB921 filter; these velocities can be derived from the filters’ central wavelengths and FWHM values). Therefore, caution is needed to look into this hypothesis. A realistic simulation is needed to evaluate this effect.

There is also a shortage of works that utilise a large sample of LAEs to suggest an opposite trend compared to our results. 
Most of the works that show LAE overdensities focus on only one particular source. For instance, \cite{Overzier2006} and \cite{Venemans2004} showed an overdensity of LAEs around a radio galaxy at $z$ = 5.2. \cite{Zheng2006}, on the other hand, looked at the 5 arcmin$^2$ regions centred at a radio-loud quasar at $z$ = 5.8 and showed that it is surrounded by fainter galaxies which are possibly LAEs. However, studies that suggest a similar trend as ours \citep[e.g.,][]{Willott2005,  Banados2013, Goto2017, Mazzucchelli2017} also focus on one bright source only. This makes our work, which utilises a very large sample of LAEs, crucial in understanding the environment of LAEs.

Our results favour the scenario wherein the overdensity of faint LAEs decreases as the central LAE's luminosity increases, and this is not entirely due to their feedback caused by their escaping photons. A similar scenario is presented \cite{Kikuta2017} where they found underdense regions of LAEs around two QSOs at $z$ $\sim$ 4.9 using the Suprime-Cam (S-Cam; \citealt{Miyazaki2002} on the Subaru Telescope). Their derived $J_{21}$ values at 3 pMpc from the two QSOs are 1.2 and 0.7, which are higher than our upper limit estimates even when we measure at 1 pMpc from our central LAEs. Although their estimates are similar to assumed values of $J_{21}$ that predict strong feedback, many factors cause uncertainty in ensuring that feedback causes the suppression of LAEs around these QSOs. These factors include uncertainties on when QSO feedback turns on relative to the formation of surrounding LAEs, short-timescale variabilities of the bright source, and complexity of heat transfer in pre-galactic clouds of LAEs due to metals and dust grains originating from nearby galaxies. Only LAEs with Ly$\alpha$ luminosity less than $10^{42}$ erg s$^{-1}$ can be suppressed by their estimated $J_{21}$ values, which are lower than the Ly$\alpha$ luminosities of our faint LAEs. Therefore, it is also difficult to show that escaping photons are the ones responsible for their results. Nevertheless, we can point out the severity of this problem by showing the same conclusion with 3 orders of magnitude larger number of sources instead of just focusing on one or two sources.

Bright LAEs prefer massive haloes \citep[e.g.,][]{Yajima2018}. More massive haloes exhibit more instances of clustering compared to less massive haloes \citep[e.g.,][]{Wetzel2007}. Therefore, we should expect to see brighter LAEs to be located in denser regions and surrounded by many galaxies. Interestingly, we observe the opposite trend. This may also imply that LAEs undergo completely different physics when it comes to suppressing their SF activities.

It is also important to take into consideration the effect of the neutral intergalactic medium (IGM) in our results since we are focusing on LAEs at $z$ $\simeq$ 5.7 and 6.6 when the epoch of reionisation is still at play. Previous works have suggested that the reionisation may have reached its end at $z$ $\simeq$ 6 \citep[e.g.,][]{Becker2001}, but other works suggest redshift values higher than 6 for the end of reionisation \citep[e.g.,][]{Wyithe2004}. The fraction of neutral hydrogen ($\Bar{x}_{\rm H I}$) increases with redshift (i.e., $\Bar{x}_{\rm H I}$ = 1 before reionisation, $\Bar{x}_{\rm H I}$ = 0 after completion of reionisation). As the fraction of neutral hydrogen increases, its impact on the escape of Ly$\alpha$ photons and therefore the observability of LAEs becomes more significant since Ly$\alpha$ photons are resonantly scattered by neutral hydrogen. Previous works \citep[e.g.,][]{Jensen2014, Matthee2015, Bosman2020} considered the effect of $\Bar{x}_{\rm H I}$ in the IGM in understanding the clustering of LAEs at high redshift. Their works showed that faint LAEs may be observed if they are inside the ionized spheres of more luminous LAEs, or when they are strongly clustered. \cite{Jensen2014} mentioned that only the most massive sources or those that are located at the densest regions will be detected in an environment with more neutral IGM. More clustered LAEs will create larger ionized regions around them, further improving Ly$\alpha$ detectability. Considering this possible scenario leads us to possibly steepen the slope, and increase the expected number of faint LAEs within the vicinity of faint central LAEs. Nevertheless, in our study, we found less faint LAEs around more luminous central LAEs. The dependence is the opposite of what is expected from the effect of neutral IGMs. This shows that (i) our results are not an artifact from the neutral IGM effect, and (ii) once we correct for the IGM effect, we might find an even stronger correlation in Fig.~\ref{fig:density_vs_luminosity}. Quantifying the effect of neutral IGM in our results will be an interesting direction to look for future work.

\section{Conclusion}
\label{sec:conclusion}

We investigated the local environment of LAEs in the \textit{SILVERRUSH} catalogue whose sources are located at $z$ $\sim$ 5.7 and $z$ $\sim$ 6.6. In contrast to previous studies which investigated QSO environments one by one, by utilising \textit{SILVERRUSH}'s unprecedentedly large sample of LAEs, we aimed to bring a statistical result on the environmental effects of LAEs. We identified bright and faint LAEs based on our magnitude cut, AB mag = 24.75. We measured the density of faint LAEs around each source within varying annulus sizes ranging from 1 pMpc to 10 pMpc. An edge correction was applied to provide a reliable estimate of number counts for those sources that are near the survey edge.

We found that the density of faint LAEs decreases with increasing bright central LAE luminosity (for $z$ $\simeq$ 5.7 sources for all distances within 10 pMpc, and for $z$ $\simeq$ 6.6 but only at distances < 1.0 pMpc from the bright LAEs), and this suppression is more efficient at smaller radial distances from the central LAE. To our knowledge, this is the first time to show such a trend for LAE environments with a 3 order of magnitude larger number of sources compared to previous works. However, our upper limit estimates of the isotropic background UV intensity due to bright LAEs show that only faint LAEs with $M_{\rm vir} < 10^{9} \ {\rm M}_\odot$ should have their SF activities weakened. However, considering the average halo mass of \textit{SILVERRUSH} LAEs, this does not fully explain the suppression of faint LAEs with $10^{9} \ {\rm M}_\odot \leq M_{\rm vir} < 10^{11} \ {\rm M}_\odot$. Our work is consistent with the scenario wherein the isotropic background UV radiation from the bright LAEs is not sufficient to explain the preference for brighter LAEs in underdense regions \citep{Kikuta2017}. 

It is possible that bright LAEs were once QSOs before but stopped their QSO activities at the time of observation due to their limited luminous phase \citep{Wyithe2005, Vanzella2011, Martini2004}. Due to many uncertainties in the properties of LAEs at higher redshift, future work is crucial to shed light on the main reason behind the suppression of faint LAEs as the luminosities of the central LAEs increase. Simulations on the effect of UV radiation on smaller halo masses as a function of halo-centric radius, and velocity dispersions and SF time scales of LAEs at very high redshift could be key directions in understanding faint LAE suppression. It is also important to study the environments of large samples of LAEs at a wider luminosity range to have a more general view of the effects of Ly$\alpha$ luminosity in the suppression of nearby faint LAEs. In addition, taking into consideration the effect of IGM in the observability of our LAE sample and increasing ionised bubble size in regions with higher LAE clustering, and investigating the number density of faint LAEs as a function of equivalent widths will give us a different perspective on how galaxy environment plays a role in the properties of LAEs at high redshift.

\section*{Acknowledgements}
We thank the anonymous referee for the insightful and constructive comments. TG and TH acknowledge the support of the National Science and Technology Council in Taiwan through grants 108-2628-M-007-004-MY3 and 110-2112-M-005-013-MY3, respectively.
TH was supported by the Centre for Informatics and Computation in Astronomy (CICA) at National Tsing Hua University (NTHU) through a grant from the Ministry of Education of the Republic of China (Taiwan).

\section*{Data Availability}
 
The \textit{SILVERRUSH} LAE catalogue is publicly available online at \url{http://cos.icrr.u-tokyo.ac.jp/rush.html}







\appendix

\section{AB magnitude to Luminosity}
\label{sec:mag_to_lum}

Here, we show how we calculate the Ly$\alpha$ luminosities of our sources given their \textit{NB816} and \textit{NB921} AB magnitudes, and vice versa.

We first calculate the flux density per unit wavelength, ${\rm F}_{\rm \nu}$, in Jy, given the AB magnitude, ${\rm m}_{\rm AB}$:

\begin{eqnarray}
    {\rm F}_\nu = 3631 \ {\rm Jy} \times 10^{\frac{-{\rm m}_{\rm AB}}{2.5}}
	\label{eq:mAB_to_flux_density_nu}
\end{eqnarray}

From ${\rm F}_{\rm \nu}$, we now calculate the flux density per unit wavelength, ${\rm F}_{\rm \lambda}$, using the following formula:

\begin{eqnarray}
    \frac{{\rm F}_\lambda}{{\rm erg} \ {\rm cm}^{-2} \ {\rm s}^{-1} \ {\rm \Angstrom}^{-1}} = \frac{\left(\frac{{\rm F}_\nu}{{\rm Jy}}\right)}{(3.34 \times 10^4)\left(\frac{\lambda}{\Angstrom}\right)^2}
	\label{eq:flux_density_nu_to_flux_density_lambda}
\end{eqnarray}

We use the values $\lambda$ = 8177 $\Angstrom$ and 9215 $\Angstrom$ for \textit{NB816} and \textit{NB921} AB magnitudes, respectively. We then use the resulting flux density per unit wavelength to calculate the flux (spectral flux density), ${\rm f}_{\rm \lambda}$, in units of ${\rm erg} \ {\rm cm}^{-2} \ {\rm s}^{-1}$:

\begin{eqnarray}
    {\rm f}_{\rm \lambda} = {\rm F}_{\rm \lambda} \Delta \lambda,
	\label{eq:flux_density_lambda_to_flux}
\end{eqnarray}

where $\Delta \lambda$ = 113 $\Angstrom$ and 135 $\Angstrom$ correspond to the full-width half-maximum (FWHM) of the \textit{NB816} and \textit{NB921} NB filters. Lastly, we calculate luminosity, $L$, in units of ${\rm erg} \ {\rm s}^{-1}$, using the formula:

\begin{eqnarray}
    L = {\rm f}_{\rm \lambda} (4\pi {\rm D}_{\rm L}^2)
	\label{eq:flux_to_lum}
\end{eqnarray}

where ${\rm D}_{\rm L}$ is the luminosity distance at the given redshift ($z$ = 5.7 and 6.6 for \textit{NB816} and \textit{NB921} AB magnitudes, respectively), in ${\rm cm}$. We note that our luminosity calculation depends on the assumption that the continuum emission has a negligible component in the NB magnitude, which is instead dominated by the Ly$\alpha$ emission.

\section{Values of densities as a function of luminosity}
\label{sec:densities}

In Tables \ref{tab:percentage_decrease1} and \ref{tab:percentage_decrease2}, we report the values plotted in Fig.~\ref{fig:density_vs_luminosity} for $z$ $\simeq$ 5.7 and $z$ $\simeq$ 6.6, respectively. The luminosity bins are centered at log $L$ (Ly$\alpha$) of central LAE = [42.93, 42.98, 43.06, 43.39] for $z$ $\simeq$ 5.7, and [43.05, 43.11, 43.18, 43.51] for $z$ $\simeq$ 6.6. These luminosity bins are chosen so that each luminosity bin has at least 50 sources, preventing large contributions of statistical error due to the low number of sources. Error bars are calculated as the standard deviation divided by the square of the number of sources in each luminosity bin.

\begin{table*}
\caption{\label{tab:percentage_decrease1}The densities and errors [in units of number counts/pMpc$^2$]} at each luminosity bin [${\rm erg} \ {\rm s}^{-1}$] and annulus size [pMpc] in Fig.~\ref{fig:density_vs_luminosity} for $z$ $\simeq$ 5.7 The corresponding AB magnitude for each luminosity bin is also presented.
\begin{center}
\begin{tabular}{c|cccc}
\hline
 Luminosity bins [${\rm erg} \ {\rm s}^{-1}$] & 42.93 & 42.98 & 43.06 & 43.39 \\ 
\hline
 Magnitude bins [AB mag] & 24.72 & 24.58 & 24.39 & 23.55 \\
 \hline
 Annulus Size (pMpc) & \multicolumn{4}{c}{densities (number counts/pMpc$^2$)}  \\ 
\cline{1-5}
r $<$ 1.0          & 0.123 $\pm$ 0.025 & 0.131 $\pm$ 0.019 & 0.105 $\pm$ 0.010 & 0.102 $\pm$ 0.016 \\
1.0 $<$ r $<$ 2.0  & 0.102 $\pm$ 0.013 & 0.100 $\pm$ 0.011 & 0.085 $\pm$ 0.009 & 0.083 $\pm$ 0.010 \\
2.0 $<$ r $<$ 3.0  & 0.096 $\pm$ 0.011 & 0.096 $\pm$ 0.009 & 0.087 $\pm$ 0.009 & 0.094 $\pm$ 0.009 \\
3.0 $<$ r $<$ 5.0  & 0.109 $\pm$ 0.008 & 0.096 $\pm$ 0.007 & 0.081 $\pm$ 0.006 & 0.087 $\pm$ 0.006 \\
5.0 $<$ r $<$ 10.0 & 0.099 $\pm$ 0.006 & 0.094 $\pm$ 0.005 & 0.079 $\pm$ 0.006 & 0.083 $\pm$ 0.005 \\
\hline
\end{tabular}
\end{center}
\end{table*}

\begin{table*}
\caption{\label{tab:percentage_decrease2}The densities and errors [in units of number counts/pMpc$^2$] at each luminosity bin [${\rm erg} \ {\rm s}^{-1}$] and annulus size [pMpc] in Fig.~\ref{fig:density_vs_luminosity} for $z$ $\simeq$ 6.6. The corresponding AB magnitude for each luminosity bin is also presented.}
\begin{center}
\begin{tabular}{c|cccc}
\hline
\hline
 Luminosity bins [${\rm erg} \ {\rm s}^{-1}$] & 43.05 & 43.11 & 43.16 & 44.51 \\ 
\hline
 Magnitude bins [AB mag] & 24.71 & 24.56 & 24.39 & 23.56 \\
 \hline
 Annulus Size (pMpc) & \multicolumn{4}{c}{densities (number counts/pMpc$^2$)}  \\ 
\cline{1-5}
r $<$ 1.0          & 0.069 $\pm$ 0.016 & 0.083 $\pm$ 0.012 & 0.059 $\pm$ 0.019 & 0.032 $\pm$ 0.012 \\ 
1.0 $<$ r $<$ 2.0  & 0.057 $\pm$ 0.009 & 0.051 $\pm$ 0.007 & 0.069 $\pm$ 0.013 & 0.056 $\pm$ 0.008 \\
2.0 $<$ r $<$ 3.0  & 0.065 $\pm$ 0.009 & 0.059 $\pm$ 0.006 & 0.064 $\pm$ 0.011 & 0.063 $\pm$ 0.007 \\
3.0 $<$ r $<$ 5.0  & 0.056 $\pm$ 0.005 & 0.056 $\pm$ 0.004 & 0.050 $\pm$ 0.007 & 0.060 $\pm$ 0.005 \\
5.0 $<$ r $<$ 10.0 & 0.060 $\pm$ 0.004 & 0.056 $\pm$ 0.003 & 0.060 $\pm$ 0.006 & 0.155 $\pm$ 0.004 \\
\hline
\end{tabular}
\end{center}
\end{table*} 




\bsp	
\label{lastpage}
\end{document}